\documentclass[twocolumn,pra,showpacs]{revtex4}
\usepackage{dcolumn,times}
\usepackage{epsfig,color,psfrag,amssymb,amsmath,amsthm,amsbsy}

\newcommand{\GPE}{Gross-Pitaevskii equation}

\newcommand{\SDE}{stochastic differential equation}

\newlength\hgt
\begin{document}

\title{Three-body recombination of ultracold Bose gases using the truncated
Wigner method}

\author{A. A. Norrie, R. J. Ballagh and C. W. Gardiner}

\affiliation{Department of Physics, University of Otago, Dunedin, New Zealand}

\author{A. S. Bradley}

\affiliation{Australian Research Council Centre of Excellence for Quantum-Atom
Optics, Department of Physics, University of Queensland, Brisbane, Australia}

\begin{abstract}
We apply the truncated Wigner method to the process of three-body recombination
in ultracold Bose gases. We find that within the validity regime of the Wigner
truncation for two-body scattering, three-body recombination can be treated
using a set of coupled stochastic differential equations that include diffusion
terms, and can be simulated using known numerical methods. As an example we
investigate the behaviour of a simple homogeneous Bose gas.
\end{abstract}

\pacs{03.75.Nt, 05.10.Gg}

\maketitle

\section{Introduction}

The dominant loss process affecting ultracold gaseous alkali metal systems is
inelastic three-body recombination
\cite{Moerdijk1996a,Wieman1997a,Dalibard1999a,Esry1999a}, a process
characterised by collisional events involving three atoms leading to the
creation of a single two-atom molecule (a dimer). The binding energy released
by the molecule formation is retained by the particles as kinetic energy.
Typically this results in the loss of all three atoms from the system as the
molecule is not trapped by any applied external potential and the energy of the
free atom is high enough to overcome any confinement barrier. Indeed it is this
process that limits the lifetime of experimentally produced alkali metal
Bose-Einstein condensates, due to the large increase in density once the
temperature is lowered past the critical point \cite{Ketterle1995a}.

In a previous paper \cite{Norrie2005b} we presented a comprehensive treatment
of the truncated Wigner approach for ultracold Bose gases including elastic
two-body interactions. In this paper we extend that treatment to include
three-body recombination events, which modifies the ensemble differential
equations describing the evolution of a single realisation of the field. These
modified differential equations are explicitly stochastic, including dynamic
noise sources arising from the action of three-body recombination on the
virtual particle background field.

To provide a demonstration of our extended formalism we examine the evolution
of a simple homogeneous system, starting from a zero-temperature state where
the particle population is initially confined to a single (condensate) mode.

\subsection{Three-body recombination in ultracold gases}

Assuming that three-body recombination is the only particle loss mechanism
affecting the system, it can be shown that the rate of change of total particle
number is \cite{Wieman1997a}
\begin{equation}
\label{general TBR number loss}
\frac{dN \left( t \right)}{dt} =
- 3 K_3 \int d{\bf x} \, g^{\left(3 \right)} \left( {\bf x},t \right)
n \left( {\bf x},t \right)^3,
\end{equation}
where $n \left( {\bf x},t \right)$ is the total number density of atoms and
$K_3$ is the three-body recombination \emph{event} rate constant. We have
assumed that all the particles involved in the recombination process are lost
from the system, hence the prefactor of 3 in Eq.~(\ref{general TBR number
loss}), which describes the \emph{number} of particles lost from the system.
The third-order normalised equiposition correlation function $g^{\left( 3
\right)} \left( {\bf x},t \right)$ measures the statistics of the field, being
unity for a fully coherent system, \emph{i.e.} a zero-temperature condensate,
and $3! = 6$ for a purely thermal system. The factor of 6 increase in the loss
rate of thermal over coherent systems for similar densities has been observed
experimentally \cite{Wieman1997a}.

\section{Truncated Wigner treatment}

\subsection{The restricted field}

As in our previous work \cite{Norrie2005b}, we describe the many-body system of
identical bosons using the Schr\"odinger picture bosonic field operator
\begin{equation}
\label{Total field operator}
\hat{\Psi} \left( {\bf x} \right) = \sum_{j} \psi_j \left( {\bf x} \right)
\hat{a}_j.
\end{equation}
Here the mode operators $\hat{a}_j$ annihilate a single boson from the $j$th
mode, and obey the commutation relations
\begin{equation}
\left[ \hat{a}_i, \hat{a}_j \right] = \left[ \hat{a}_i^{\dagger},
\hat{a}_j^{\dagger} \right] = 0, \hspace{0.5cm}
\left[ \hat{a}_i, \hat{a}_j^{\dagger} \right] = \delta_{i,j},
\end{equation}
while the coordinate space functions $\psi_j \left( {\bf x} \right)$ form an
infinite orthonormal basis set where
\begin{equation}
\left[ -\frac{\hbar^2 \nabla^2}{2m} + U_{\rm ext} \left( {\bf x} \right) \right]
\psi_j \left( {\bf x} \right) = \hbar \omega_j \psi_j \left( {\bf x} \right),
\end{equation}
where $U_{\rm ext} \left( {\bf x} \right)$ is the applied external potential.

We now divide mode space into two subspaces, a low-energy space ($L$)
consisting of all those modes whose eigenenergies are less than the boundary
energy $\varepsilon_{\rm cut}$, and a high-energy space ($H$) that includes all
remaining modes. For this work our interest lies with the dynamics of the
low-energy subspace. Using these subspaces we define the field operators
\begin{subequations}
\begin{eqnarray}
\hat{\Psi}_{\mathcal{P}} \left( {\bf x} \right) & \equiv &
\sum_{j \in L} \psi_j \left( {\bf x} \right) \hat{a}_j \\
\hat{\Psi}_{\mathcal{Q}} \left( {\bf x} \right) & \equiv &
\sum_{j \in H} \psi_j \left( {\bf x} \right) \hat{a}_j,
\end{eqnarray}
\end{subequations}
Of most importance for this paper is the low-energy \emph{restricted basis
field operator} $\hat{\Psi}_{\mathcal{P}} \left( {\bf x} \right)$, which can be
obtained from the total field operator
\begin{equation}
\hat{\Psi} \left( {\bf x} \right) \equiv \sum_{j \in L, H} \psi_j \left(
{\bf x} \right) \hat{a}_j,
\end{equation}
using the projector
\begin{equation}
\label{Projector defined}
\mathcal{P} \equiv \sum_{j \in L} \left| j \right \rangle \left \langle j
\right|
\end{equation}
as $\hat{\Psi}_{\mathcal{P}} \left( {\bf x} \right) = \mathcal{P} \left[
\hat{\Psi} \left( {\bf x} \right) \right]$. The restricted basis field operator
obeys the commutation relations
\begin{eqnarray}
\left[ \hat{\Psi}_{\mathcal{P}} \left( {\bf x} \right),
\hat{\Psi}_{\mathcal{P}} \left( {\bf x}' \right) \right] & = &
\left[ \hat{\Psi}_{\mathcal{P}}^{\dagger} \left( {\bf x} \right),
\hat{\Psi}_{\mathcal{P}}^{\dagger} \left( {\bf x}' \right) \right] = 0
\nonumber \\
\left[ \hat{\Psi}_{\mathcal{P}} \left( {\bf x} \right),
\hat{\Psi}_{\mathcal{P}}^{\dagger} \left( {\bf x}' \right) \right] & = &
\delta_{\mathcal{P}} \left( {\bf x},{\bf x}' \right), \hspace{1cm}
\end{eqnarray}
where the \emph{restricted delta function} is defined by
\begin{equation}
\label{Restricted delta function}
\delta_{\mathcal{P}} \left( {\bf x},{\bf x}' \right) \equiv
\sum_{j \in L} \psi_j^* \left( {\bf x}' \right)
\psi_j \left( {\bf x} \right).
\end{equation}
The conjugate projector $\mathcal{Q}$ can be obtained using the complementarity
relation $\mathcal{P} + \mathcal{Q} = 1$.

\subsection{Master equation}

Our previous paper \cite{Norrie2005b} assumed that only two atoms participate
in any single scattering event. In this way the particle interactions are
described using a simple $s$-wave contact potential as an approximation to the
full two-body T-matrix. Obviously such a description does not include
three-body scattering events. Full theoretical treatments of three-body
scattering including all possible collisional channels are extremely
complicated, and we do not attempt such an approach here. Instead we adopt a
quantum-optical approach, starting from a phenomenologically appropriate
Hamiltonian including inelastic three-body recombination events, to which we
apply the truncated Wigner method.

We assume that within the dilute limit the characteristic range of the
three-body recombination potential $U_{\rm TBR} \left( {\bf x}_1, {\bf x}_2,
{\bf x}_3 \right)$ is much smaller than the average interparticle spacing.
Thus, following thematically the approach for pairwise scattering, we replace
this scattering potential by an effective zero-range \emph{three-body}
scattering T-operator, whose interaction strength is essentially a free
parameter that will be chosen to satisfy experimentally observed loss rates.
Within this approach then, in order to include the effects of three-body
recombination the Schr\"odinger picture effective Hamiltonian is modified to
include the term \cite{Jack2002a}
\begin{widetext}
\begin{equation}
\label{TBR Hamiltonian}
\hat{H}_{\rm eff}^{\rm \left( TBR \right)} = \kappa_{\rm TBR}
\int d{\bf x} \left[ \hat{\Psi}_{\mathcal{Q}}^{\dagger} \left( {\bf x} \right)
\hat{\Xi}^{\dagger} \left( {\bf x} \right)
\left( \hat{\Psi}_{\mathcal{P}} \left( {\bf x} \right) \right)^3 +
\left( \hat{\Psi}_{\mathcal{P}}^{\dagger} \left( {\bf x} \right) \right)^3
\hat{\Psi}_{\mathcal{Q}} \left( {\bf x} \right)
\hat{\Xi} \left( {\bf x} \right) \right].
\end{equation}
\end{widetext}
where the molecule field operator $\hat{\Xi} \left( {\bf x} \right)$
annihilates a dimer from the field and $\kappa_{\rm TBR}$ is a measure of the
energy associated with the three-body process. We have assumed in formulating
Eq.~(\ref{TBR Hamiltonian}) that the binding energy associated with the
molecule formation is large enough that the unpaired atom generated by a
recombination event is created within the high-energy subspace $H$, rather than
the low-energy (system) subspace $L$, and is described by the high-energy field
operator $\hat{\Psi}_{\mathcal{Q}} \left( {\bf x} \right)$.

Jack \cite{Jack2002a} has considered this partial Hamiltonian, and has shown
that by eliminating both the molecular and high-energy atomic fields from the
evolution using a standard interaction picture approach for initially uncoupled
fields \cite{Walls1994a}, one obtains the master equation term
\begin{widetext}
\begin{eqnarray}
\label{TBR master equation}
\frac{d \rho \left( t \right)}{dt}^{\rm \left( TBR \right)}
& = & \frac{\gamma}{6} \int d{\bf x} \left[
2 \left( \hat{\Psi}_{\mathcal{P}} \left( {\bf x} \right) \right)^3
\rho \left( t \right)
\left( \hat{\Psi}_{\mathcal{P}}^{\dagger} \left( {\bf x} \right) \right)^3
- \left( \hat{\Psi}_{\mathcal{P}}^{\dagger} \left( {\bf x} \right) \right)^3
\left( \hat{\Psi}_{\mathcal{P}} \left( {\bf x} \right) \right)^3
\rho \left( t \right) \right. \nonumber \\
&& \left. \mbox{} - \rho \left( t \right)
\left( \hat{\Psi}_{\mathcal{P}}^{\dagger} \left( {\bf x} \right) \right)^3
\left( \hat{\Psi}_{\mathcal{P}} \left( {\bf x} \right) \right)^3
\right],
\end{eqnarray}
\end{widetext}
for the low-energy atomic subspace (system) density operator $\rho \left( t
\right)$. The quantity $\gamma$ governs the rate of recombination events, and
its relationship to $K_3$ we consider later. In arriving at this master
equation term it has been assumed that the output products of the recombination
events immediately exit the coordinate space region containing the system, such
that they play no further role in the evolution. To describe the full master
equation for the system density operator one combines Eq.~(\ref{TBR master
equation}) with the von Neumann equation calculated in \cite{Norrie2005b}.

\subsection{Functional Wigner function correspondences}

The master equation term given by Eq.~(\ref{TBR master equation}) can be used
to calculate the evolution of the corresponding multimode Wigner function $W
\left( \left\{ \alpha_j,\alpha_j^* \right\},t \right)$ using appropriate
\emph{operator correspondences} \cite{Gardiner2000a}. However, rather than
using the mode operator correspondences that were used in \cite{Norrie2005b},
here we perform this step using \emph{functional operator correspondences}.

We define, similar to the restricted basis field operator
$\hat{\Psi}_{\mathcal{P}} \left( {\bf x} \right)$, the restricted basis
wavefunctions
\begin{subequations}
\label{Restricted basis wavefunctions}
\begin{eqnarray}
\Psi_{\mathcal{P}} \left( {\bf x} \right) & \equiv &
\sum_{j \in L} \psi_j \left( {\bf x} \right) \alpha_j \\
\Psi^*_{\mathcal{P}} \left( {\bf x} \right) & \equiv &
\sum_{j \in L} \psi_j^* \left( {\bf x} \right) \alpha_j^*
\end{eqnarray}
\end{subequations}
and the related \emph{functional derivatives}
\begin{subequations}
\label{Functional derivative operators}
\begin{eqnarray}
\frac{\delta}{\delta \Psi_{\mathcal{P}} \left( {\bf x} \right)} & \equiv &
\sum_{j \in L} \psi_j^* \left( {\bf x} \right)
\frac{\partial}{\partial \alpha_j} \\
\frac{\delta}{\delta \Psi^*_{\mathcal{P}} \left( {\bf x} \right)} & \equiv &
\sum_{j \in L} \psi_j \left( {\bf x} \right)
\frac{\partial}{\partial \alpha_j^*}.
\end{eqnarray}
\end{subequations}

Using these definitions together with the Wigner function mode operator
correpondences \cite{Gardiner2000a}, we find that the actions of the restricted
basis field operator on the system density operator $\rho \left( t \right)$ can
be expressed as actions on the corresponding Wigner function using
\begin{subequations}
\label{functional operator correspondences}
\begin{eqnarray}
\hat{\Psi}_{\mathcal{P}} \left( {\bf x} \right)
\rho \left( t \right) & \leftrightarrow &
\left( \Psi_{\mathcal{P}} \left( {\bf x} \right) + \frac{1}{2}
\frac{\delta}{\delta \Psi^*_{\mathcal{P}} \left( {\bf x} \right)} \right)
W \left( t \right) \\
\hat{\Psi}^{\dagger}_{\mathcal{P}} \left( {\bf x} \right)
\rho \left( t \right) & \leftrightarrow &
\left( \Psi^*_{\mathcal{P}} \left( {\bf x} \right) - \frac{1}{2}
\frac{\delta}{\delta \Psi_{\mathcal{P}} \left( {\bf x} \right)} \right)
W \left( t \right) \\
\rho \left( t \right)
\hat{\Psi}_{\mathcal{P}} \left( {\bf x} \right) & \leftrightarrow &
\left( \Psi_{\mathcal{P}} \left( {\bf x} \right) - \frac{1}{2}
\frac{\delta}{\delta \Psi^*_{\mathcal{P}} \left( {\bf x} \right)} \right)
W \left( t \right) \\
\rho \left( t \right)
\hat{\Psi}^{\dagger}_{\mathcal{P}} \left( {\bf x} \right) & \leftrightarrow &
\left( \Psi^*_{\mathcal{P}} \left( {\bf x} \right) + \frac{1}{2}
\frac{\delta}{\delta \Psi_{\mathcal{P}} \left( {\bf x} \right)} \right)
W \left( t \right).
\end{eqnarray}
\end{subequations}
Such \emph{functional Wigner function operator correspondences} have been
previously used by Steel \emph{et al.} \cite{Steel1998a}.

\subsection{Wigner function evolution}

Applying the functional Wigner function operator correspondences,
Eq.~(\ref{functional operator correspondences}), to the master equation term
describing three-body recombination, Eq.~(\ref{TBR master equation}), we
obtain, after some manipulation, the Wigner function evolution term
\begin{widetext}
\begin{eqnarray}
\label{Wigner function evolution TBR full}
\frac{\partial W}{\partial t}^{\rm \left( TBR \right)} \hspace{-0.5cm} & = &
\frac{\gamma}{6} \int d{\bf x} \left[
\left( \frac{\delta}{\delta \Psi_{\mathcal{P}}} \Psi_{\mathcal{P}} +
\frac{\delta}{\delta \Psi_{\mathcal{P}}^*} \Psi_{\mathcal{P}}^* \right) \right.
\left( 3 \left| \Psi_{\mathcal{P}} \right|^4 - 9 \left| \Psi_{\mathcal{P}}
\right|^2 \delta_{\mathcal{P}} \left( {\bf x},{\bf x} \right) + \frac{9}{2}
\delta_{\mathcal{P}} \left( {\bf x},{\bf x} \right)^2 \right)
\nonumber \\ && \mbox{}
+ \frac{\delta^2}{\delta \Psi_{\mathcal{P}} \delta \Psi_{\mathcal{P}}^*} \left(
9 \left| \Psi_{\mathcal{P}} \right|^4 - 18 \left| \Psi_{\mathcal{P}} \right|^2
\delta_{\mathcal{P}} \left( {\bf x},{\bf x} \right) + \frac{9}{2}
\delta_{\mathcal{P}} \left( {\bf x},{\bf x} \right)^2 \right)
+ \frac{1}{4} \left( \frac{\delta^3}{\delta \Psi_{\mathcal{P}}^3}
\Psi_{\mathcal{P}}^3 + \frac{\delta^3}{\delta \Psi_{\mathcal{P}}^{*3}}
\Psi_{\mathcal{P}}^{*3} \right)
\nonumber \\ && \mbox{}
+ \frac{9}{4} \left( \frac{\delta^3}{\delta \Psi_{\mathcal{P}}^2 \delta
\Psi_{\mathcal{P}}^*} \Psi_{\mathcal{P}} + \frac{\delta^3}{\delta
\Psi_{\mathcal{P}} \delta \Psi_{\mathcal{P}}^{*2}} \Psi_{\mathcal{P}}^* \right)
\left( \left| \Psi_{\mathcal{P}} \right|^2 - \delta_{\mathcal{P}} \left( {\bf
x},{\bf x} \right) \right)
+ \frac{3}{4} \left( \frac{\delta^4}{\delta \Psi_{\mathcal{P}}^3 \delta
\Psi_{\mathcal{P}}^*} \Psi_{\mathcal{P}}^2 + \frac{\delta^4}{\delta
\Psi_{\mathcal{P}} \delta \Psi_{\mathcal{P}}^{*3}} \Psi_{\mathcal{P}}^{*2}
\right)
\nonumber \\ && \left. \mbox{}
+ \frac{3}{16} \left( \frac{\delta^5}{\delta \Psi_{\mathcal{P}}^3 \delta
\Psi_{\mathcal{P}}^{*2}} \Psi_{\mathcal{P}} + \frac{\delta^5}{\delta
\Psi_{\mathcal{P}}^2 \delta \Psi_{\mathcal{P}}^{*3}} \Psi_{\mathcal{P}}^*
\right)
+ \frac{1}{16} \frac{\delta^6}{\delta \Psi_{\mathcal{P}}^3 \delta
\Psi_{\mathcal{P}}^{*3}} \right] W.
\end{eqnarray}
\end{widetext}
Eq.~(\ref{Wigner function evolution TBR full}) is a rather complex equation of
motion, including derivative terms up to sixth order. However, we can only
write differential equations describing the evolution of a single ensemble
member for Wigner function evolutions containing derivative terms up to second
order. Thus to proceed we must truncate the higher order terms in
Eq.~(\ref{Wigner function evolution TBR full}), a process that is also required
for the pairwise scattering \cite{Norrie2005b}.

\subsubsection{Wigner truncation}

To justify the truncation of the higher order terms in the Wigner function
evolution, we follow a similar method to that given in \cite{Norrie2005b}. Let
us assume that at some time $\tau$ the Wigner function of the (low-energy)
system has the factorisable form
\begin{equation}
\label{Full Wigner factored}
W \left( \left\{ \alpha_j, \alpha_j^* \right\}, \tau \right) =
\prod_{j \in L} \frac{\Gamma_j}{\pi}
\exp \left[ - \Gamma_j \left| \alpha_j - \alpha_{j_0} \right|^2 \right].
\end{equation}
Here $\alpha_{j_0} \equiv \left \langle \hat{a}_j \right \rangle$ is the
expectation value amplitude of the $j$th mode, and $\Gamma_j$ is proportional
to the inverse width of the Wigner function for that mode. This type of
function describes both coherent (where $\Gamma_j = 2$) and thermally
distributed modes, but does not describe number states or other more exotic
states. The factorisability of this Wigner function indicates that number
fluctuations between disparate modes are uncorrelated.

Evaluating the Wigner function evolution given by Eq.~(\ref{Wigner function
evolution TBR full}) using the Wigner function given by Eq.~(\ref{Full Wigner
factored}) returns a rather complicated expression, which we give in full in
the appendix, Eq.~(\ref{TBR Wigner evolution evaluated}). Essentially we find
that for increasing order in $\delta / \delta \Psi_{\mathcal{P}}$, the leading
order term in $\left| \Psi_{\mathcal{P}} \right|$ decreases. In performing the
Wigner truncation for the two-body scattering in \cite{Norrie2005b}, we
required that in the coordinate space regions of high real particle density
that $n \left( {\bf x} \right) \gg \delta_{\mathcal{P}} \left( {\bf x,x}
\right)$. Given that $\left| \Psi_{\mathcal{P}} \right|^2 \sim n$ and that all
remaining terms scale as $\delta_{\mathcal{P}}$, and assuming that there is
significant real particle density in the regions where three-body recombination
is important compared to the local density of modefunctions, only the first few
terms in $\left| \Psi_{\mathcal{P}} \right|$ are important. By keeping only
those terms of 4th and 5th order in $\left| \Psi_{\mathcal{P}} \right|$ we find
that, within the same validity regime for the two-body elastic scattering, the
Wigner function equation of motion can be accurately described by
\begin{widetext}
\begin{equation}
\frac{\partial W}{\partial t}^{\rm \left( TBR \right)} \approx \frac{\gamma}{2}
\int d{\bf x} \left( \frac{\delta}{\delta \Psi_{\mathcal{P}}} \Psi_{\mathcal{P}}
+ \frac{\delta}{\delta \Psi_{\mathcal{P}}^*} \Psi_{\mathcal{P}}^*
+ 3 \frac{\delta^2}{\delta \Psi_{\mathcal{P}} \delta \Psi_{\mathcal{P}}^*}
\right) \left| \Psi_{\mathcal{P}} \right|^4 W.
\end{equation}
\end{widetext}
While it is possible to directly convert this equation of motion into a set of
coupled differential equations, the functional nature of the derivative
operators can obscure some of the details. Instead we choose to perform the
conversion using an explicit mode representation. Using our definitions of the
functional derivatives, Eq.~(\ref{Functional derivative operators}), we find
that the Wigner function evolution due to three-body recombination can be
expressed as
\begin{widetext}
\begin{eqnarray}
\label{Truncated FPE modes}
\frac{\partial W}{\partial t}^{\rm \left( TBR \right)} & = &
\sum_{j \in L} \left[ \frac{\partial}{\partial \alpha_j}
\frac{\gamma}{2} \int d{\bf x}\, \psi_j^* \left| \Psi_{\mathcal{P}} \right|^4
\Psi_{\mathcal{P}} + \frac{\partial}{\partial \alpha_j^*} \frac{\gamma}{2} \int
d{\bf x} \, \psi_j \left| \Psi_{\mathcal{P}} \right|^4 \Psi_{\mathcal{P}}^*
\right] W \nonumber \\
&& \mbox{} + \frac{1}{2} \sum_{ij \in L}
\frac{\partial^2}{\partial \alpha_j \partial \alpha_i^*} 3 \gamma
\int d{\bf x} \, \psi_j^* \left| \Psi_{\mathcal{P}} \right|^4 \psi_i \, W.
\end{eqnarray}
\end{widetext}

\subsection{Stochastic differential equations}

It is important to remember when converting Eq.~(\ref{Truncated FPE modes}) to
its equivalent differential equations that we have two sets of independent
variables, $\left\{ \alpha_j \right\}$ and $ \left\{ \alpha_j^* \right\}$. Thus
while the drift terms are straightforward, and we find by using the relations
given in \cite{Gardiner2003b} for the Ito calculus that
\begin{eqnarray}
\label{TBR drift evolution}
A_j & = & - \frac{\gamma}{2} \int d{\bf x} \, \psi_j^*
\left| \Psi_{\mathcal{P}} \right|^4 \Psi_{\mathcal{P}} \nonumber \\
A_{j^*} & = & - \frac{\gamma}{2} \int d{\bf x} \, \psi_j
\left| \Psi_{\mathcal{P}} \right|^4 \Psi_{\mathcal{P}}^*,
\end{eqnarray}
where $A_{j^*} = A_j^*$ as required, the diffusion terms are not so easily
obtained.

To obtain the terms in the stochastic differential equations corresponding to
the diffusion terms in Eq.~(\ref{Truncated FPE modes}), we first find it
necessary to rewrite the coefficient of that diffusive part as
\begin{widetext}
\begin{eqnarray}
\int d{\bf x} \, \psi_j^* \left| \Psi_{\mathcal{P}} \right|^4 \psi_i & = &
\int d{\bf x} \int d{\bf x}' \, \psi_j^* \left| \Psi_{\mathcal{P}} \right|^2
\delta \left( {\bf x - x}' \right) \left| \Psi_{\mathcal{P}}' \right|^2 \psi_i'
\\
& = & \int d{\bf x} \int d{\bf x}' \, \psi_j^*
\left| \Psi_{\mathcal{P}} \right|^2 \sum_{k \in L,H} \psi_k
\left( \psi_k' \right)^* \left| \Psi_{\mathcal{P}}' \right|^2 \psi_i' \\
& = & \sum_{k \in L,H} \int d{\bf x} \, \psi_j^* \left|
\Psi_{\mathcal{P}} \right|^2 \psi_k \int d{\bf x}' \, \left( \psi_k' \right)^*
\left| \Psi_{\mathcal{P}}' \right|^2 \psi_i',
\label{diffusion coefficient split 3}
\end{eqnarray}
\end{widetext}
where it is important to note that the summation over the index $k$ runs over
the complete mode space $L \oplus H$. Including basis modes that are not part
of the system subspace into the formalism in this way may appear to be cause
for concern, as the master equation term from which we are working,
Eq.~(\ref{TBR master equation}) contains no reference to these high-energy
states. However, we note that we are free to choose any set of ensemble
differential equations that can be shown to be mathematically equivalent to the
Fokker-Planck equation \cite{Gardiner2003b}, such that our inclusion of the
high-energy modes in Eq.~(\ref{diffusion coefficient split 3}) is certainly
mathematically accurate.

It can be shown, either by rewriting Eq.~(\ref{Truncated FPE modes}) in terms
of explicitly real quantities (including the mode amplitude quadratures) and
directly using the conversion relations given in \cite{Gardiner2000a}, or by
working backwards using complex Ito calculus, that the ensemble differential
equations corresponding to the diffusive part of the Wigner function evolution
are given by 
\begin{equation}
\label{mode evolution diffusive}
d \alpha_j^{\rm \left( diff \right)} = \sqrt{\frac{3\gamma}{2}} \int d{\bf x} \,
\psi_j^* \left| \Psi_{\mathcal{P}} \right|^2 \sum_{k \in L,H} \psi_k \, dW_k,
\end{equation}
for all those modes $j \in L$. Here the complex Wiener processes $dW_k \left( t
\right)$ obey the relations
\begin{subequations}
\begin{eqnarray}
\left \langle dW_k \left( t \right) \right \rangle & = & 0 \\
\left \langle dW_k \left( t \right) dW_l \left( t \right)
\right \rangle & = & 0 \\
\left \langle dW_k \left( t \right) dW_l^* \left( t \right)
\right \rangle & = & \delta_{k,l} \, dt.
\end{eqnarray}
\end{subequations}
In fact, given the local nature of the recombination process in coordinate
space, a more useful form of the \emph{total} Wiener process is given by
\begin{equation}
\label{spatial Wiener process}
dW \left( {\bf x},t \right) \equiv \sum_{k \in L,H}
\psi_k \left( {\bf x} \right) dW_k \left( t \right),
\end{equation}
which can be straightforwardly shown to obey
\begin{subequations}
\label{spatial Wiener properties}
\begin{eqnarray}
\left \langle dW \left( {\bf x},t \right) \right \rangle & = & 0 \\
\left \langle dW \left( {\bf x},t \right) dW \left( {\bf x}',t \right)
\right \rangle & = & 0 \\
\left \langle dW \left( {\bf x},t \right) dW^* \left( {\bf x}',t \right)
\right \rangle & = & \delta \left( {\bf x - x}' \right) dt.
\end{eqnarray}
\end{subequations}

Inserting the spatial Wiener process $dW$ as appropriate into the diffusive
mode evolution given by Eq.~(\ref{mode evolution diffusive}), using the drift
mode evolutions of Eq.~(\ref{TBR drift evolution}) and including the evolution
in the absence of three-body recombination given in \cite{Norrie2005b}, gives
the total evolution of the low-energy system mode amplitudes as
\begin{widetext}
\begin{equation}
\label{SDE total with TBR hybrid 1}
d\alpha_j = -i \omega_j \alpha_j \, dt
+ \int d{\bf x} \, \psi_j^* \left\{ - \left[
\frac{i}{\hbar} U_0 \left| \Psi_{\mathcal{P}} \right|^2
+ \frac{\gamma}{2} \left| \Psi_{\mathcal{P}} \right|^4 \right]
\Psi_{\mathcal{P}} \, dt
+ \sqrt{\frac{3\gamma}{2}} \left| \Psi_{\mathcal{P}} \right|^2
dW \left( {\bf x},t \right) \right\}.
\end{equation}
Using the definition of the system wavefunction given by Eq.~(\ref{Restricted
basis wavefunctions}) we find that the corresponding evolution of the
coordinate space field is
\begin{equation}
\label{total wavefunction evolution final 1}
d \Psi_{\mathcal{P}} =
-\frac{i}{\hbar} \left[ -\frac{\hbar^2 \nabla^2}{2m} + U_{\rm ext} \right]
\Psi_{\mathcal{P}} \, dt
+ \mathcal{P} \left\{ -\left[ \frac{i}{\hbar} U_0 \left| \Psi_{\mathcal{P}}
\right|^2 + \frac{\gamma}{2} \left| \Psi_{\mathcal{P}} \right|^4 \right]
\Psi_{\mathcal{P}} \, dt + \sqrt{\frac{3 \gamma}{2}} \left| \Psi_{\mathcal{P}}
\right|^2 dW \left( {\bf x},t \right) \right\},
\end{equation}
\end{widetext}
where we have recognised the low-energy projector $\mathcal{P}$,
Eq.~(\ref{Projector defined}).

\subsubsection{Rate of population change}

The total (real) particle population of the field is defined as
\begin{equation}
N \equiv \sum_{j \in L} \left \langle \hat{N_j} \right \rangle
= \sum_{j \in L} \left \langle \hat{a}_j \hat{a} \right \rangle.
\end{equation}
Using the correspondence of moments of the Wigner function to symmetrically
ordered products of quantum operators \cite{Gardiner2000a}, we find that $N$
can be calculated using Wigner function averages as
\begin{equation}
\label{Wigner field population}
N = \sum_{j \in L} \left \langle \left| \alpha_j \right|^2 - \frac{1}{2} \right
\rangle_W.
\end{equation}
Using this result, we find the rate of change of particle number for a single
trajectory to be given by Ito's formula \cite{Gardiner2003b}
\begin{equation}
\frac{dN}{dt} = \lim_{dt \rightarrow 0} \frac{1}{dt} \sum_{j \in L}
\left( d\alpha_j^* \alpha_j + \alpha_j^* d\alpha_j +
d\alpha_j^* d\alpha_j \right).
\end{equation}
The terms $\left\{ d\alpha_j^* d\alpha_j \right\}$ are included here because,
unlike ordinary deterministic calculus, the presence of the Wiener processes in
Eq.~(\ref{SDE total with TBR hybrid 1}) give these terms a non-zero value in
the limit $dt \rightarrow 0$.

Taking expectation values, using the properties of the spatially-dependent
Wiener process, Eq.~(\ref{spatial Wiener properties}), we find the ensemble
averaged rate of normalisation change to be
\begin{equation}
\label{expected rate of population change}
\left \langle \frac{dN}{dt} \right \rangle =
-\gamma \int d{\bf x} \left[
\left \langle \left| \Psi_{\mathcal{P}} \right|^6 \right \rangle_W -
\frac{3}{2} \delta_{\mathcal{P}}
\left \langle \left| \Psi_{\mathcal{P}} \right|^4 \right \rangle_W \right],
\end{equation}
where we have written $\delta_{\mathcal{P}}$ for $\delta_{\mathcal{P}} \left(
{\bf x,x} \right)$ for compactness, as we also do below. It may appear from
Eq.~(\ref{expected rate of population change}) that our truncated Wigner
treatment of three-body recombination introduces a small correction to the rate
of particle loss, apparently creating particles (in the average). However, a
clearer understanding can be obtained by expressing the moments of the Wigner
function as physically significant quantities.

Using the properties of the Wigner function moments we find, for example, that
\begin{eqnarray}
\hspace{-0.4cm}
\left \langle \left| \Psi_{\mathcal{P}} \right|^4 \right \rangle_W
\hspace{-0.3cm} & = &
\left \langle \left(\hat{\Psi}_{\mathcal{P}}^{\dagger} \right)^2
\hat{\Psi}_{\mathcal{P}}^2 \right \rangle
+ 2 \left \langle \hat{\Psi}_{\mathcal{P}}^{\dagger} \hat{\Psi}_{\mathcal{P}}
\right \rangle \delta_{\mathcal{P}} + \frac{1}{2} \delta_{\mathcal{P}}^2 \\
& = &
g^{\left( 2 \right)} n^2 + 2 n \delta_{\mathcal{P}}
+ \frac{1}{2} \delta_{\mathcal{P}},
\end{eqnarray}
where we have again suppressed the spatial dependences. Replacing the Wigner
function moments in Eq.~(\ref{expected rate of population change}) in this way
returns
\begin{equation}
\label{expected rate of population change 2}
\left \langle \frac{dN}{dt} \right \rangle =
-\gamma \int d{\bf x} \left[ g^{\left( 3 \right)} n^3
+ 3 g^{\left( 2 \right)} n^2 \delta_{\mathcal{P}}
+ \frac{3}{2} n \delta_{\mathcal{P}}^2 \right]
\end{equation}
Thus, rather than reducing the rate of particle loss, the truncated Wigner
treatment leads to an increased rate of particle loss. However, given that we
have required that $n \gg \delta_{\mathcal{P}}$ to perform the Wigner
truncation, this correction should be small. Note that Eq.~(\ref{expected rate
of population change 2}) also shows that particle loss only occurs in those
regions where there is real particle density, such that those coordinate space
regions solely occupied by virtual particles will exhibit zero particle loss.

Comparing Eq.~(\ref{expected rate of population change 2}) to Eq.~(\ref{general
TBR number loss}) shows that $\gamma = 3K_3$. Thus while $K_3$ is the rate
constant for three-body recombination \emph{events}, $\gamma$ is the
\emph{number} loss rate constant.

It is worthwhile discussing a possible point of confusion when using these
classical field methods. As the field for a single trajectory is represented by
a single wavefunction, it could be considered that the field is therefore
uniformly coherent at all points. In such a case those behaviours that depend
upon the statistics of the field, such as three-body recombination, would be
improperly treated. However, this view is incorrect, as such statistics only
obtain physical meaning when considering ensembles of trajectories. As an
example, while direct inclusion of the statistics is relevant when considering
three-body recombination using the \emph{mean} (either spatial or temporal)
particle density, in those regions where the system exhibits thermal
statistics, the trajectory wavefunction will exhibit density fluctuations both
spatially and temporally. Thus the (spatial and temporal) mean of $\left|
\Psi_{\mathcal{P}} \right|^6$ will be larger than the mean density cubed,
leading to the increased rate of loss observed experimentally. Indeed, given
that a single trajectory is entirely analogous to a single experimental run,
the fact that here the wavefunction contains the full behaviour of the field
is, if not obvious, at least eminently reasonable. This result also provides
for the factor of two increase in nonlinear interaction strength between the
condensate and the thermal particles due to the exchange energy
\cite{Stringari2003a}.

\subsection{Plane wave basis}

While our formalism is applicable to any orthonormal single-particle basis
$\left\{ \psi_j \left( {\bf x} \right) \right\}$, the most useful set of
mode-functions for many situations, including the simple system we consider in
this paper, is the plane-wave modes. For this basis the modes are eigenstates
of the curvature operator only, such that $U_{\rm ext} = 0$ and $\hbar \omega_j
= \hbar^2 k_j^2 / 2m$ where $k_j \equiv \left| {\bf k}_j \right|$, and
\begin{equation}
\psi_j \left( {\bf x} \right) = \frac{1}{\sqrt{V}} e^{i {\bf k}_j \cdot {\bf
x}}.
\end{equation}
Within a periodically bounded volume of extent $V = L_x \times L_y \times L_z$
the orthonormal plane-wave modes are arranged in momentum space such that
\begin{equation}
\label{plane wave momenta}
{\bf k}_j = \frac{2\pi m_j}{L_x} \hat{\bf k}_x
+ \frac{2\pi n_j}{L_x} \hat{\bf k}_x \frac{2\pi p_j}{L_x} \hat{\bf k}_x,
\end{equation}
where $m_j$, $n_j$ and $p_j$ are integers. Using this plane-wave basis, the
energy cutoff that defines the low-energy mode subspace becomes a spherical
cutoff in momentum space, with the boundary defined by $\hbar k_{\rm cut} =
\sqrt{2m \varepsilon_{\rm cut}}$.

\section{Numerical simulations}

To demonstrate our truncated Wigner treatment of three-body recombination we
have numerically simulated a (relatively) simple zero-temperature homogeneous
gas of $\left|F,m_F \right \rangle = \left| 1,-1 \right \rangle$ $^{23}$Na
atoms. We describe the system using a set of plane-wave modes, with the initial
real particle population confined to the ground (${\bf k}_j = 0$) mode. 

Determination of the three-body recombination event rate constant $K_3$ for
various alkali metals has been performed both theoretically
\cite{Moerdijk1996a,Fedichev1996a,Esry1999a} and experimentally
\cite{Wieman1997a,Ketterle1998a,Dalibard1999a,Ketterle1999a}. For this paper we
take as a best estimate of the relevant $K_3$ the value measured at MIT
\cite{Ketterle1998a} for a fully condensed gas
\begin{equation}
\label{TBR: K3 use}
K_3 = 3.7 \times 10^{-31} \hspace{0.1cm} {\rm cm}^6{\rm s}^{-1}.
\end{equation}
In that work it was reported that optical confinement was used to produce a
rather large particle density of $^{23}$Na of $3\times10^{15}$~cm$^{-3}$ at the
centre of the trap. Thus, given that the rate of particle loss scales as $n^3$
and the correction due to the dynamic noise sources as $n^2$, such a large
particle density should provide information on a parameter regime where
three-body recombination is significant.

We use a simulation volume of $V = \left( 4.7 \mu{\rm m} \right)^3$, such that
to achieve an initial (uniform) density of $n = 3\times10^{15}$~cm$^{-3}$ we
use $N_0 \left( t = 0 \right) = 3.06 \times 10^5$. The mode spacing (in
velocity space) along each of the cartesian directions, Eq.~(\ref{plane wave
momenta}), is determined by the volume to be 3.7~mms$^{-1}$. To characterise
the strength of the interactions, we use $a = 2.75$~nm.

We have performed simulations using two distinct boundaries to the low-energy
subspace: $v_{\rm cut} = 44.3$~mms$^{-1}$, for which the number of modes $M =
7.2 \times 10^3$; and $v_{\rm cut} = 59.1$~mms$^{-1}$, for which $M = 1.7
\times 10^4$. In both cases the number of modes is significantly less than the
number of real particles, and we therefore expect that both cutoffs will return
valid results.

\subsection{Initial states}

For our zero-temperature homogeneous system, the appropriate initial state for
a single trajectory is described by
\begin{equation}
\label{trajectory initial state}
\alpha_0 \left( 0 \right) = \sqrt{N_0} + \frac{1}{2} \left[ A_0 + i B_0
\right] \hspace{0.5cm}
\alpha_{j \neq 0} \left( 0 \right) = \frac{1}{2} \left[ A_j + i B_j \right].
\end{equation}
Here $A_j$ and $B_j$ are Gaussian random variables of zero mean and unit
variance, such that
\begin{eqnarray}
\left \langle A_j \right \rangle & = & \left \langle B_j \right \rangle = \left
\langle A_i B_j \right \rangle = 0 \nonumber \\
\left \langle A_i A_j \right \rangle & = & \left \langle B_i B_j \right \rangle
= 1.
\end{eqnarray}
This initial state, Eq.~(\ref{trajectory initial state}) satisfies the assumed
Wigner function used to justify the Wigner truncation, Eq.~(\ref{Full Wigner
factored}), with $\Gamma_j = 2$ for all $j$, $\alpha_{0_0} = \sqrt{N_0}$ and
$\alpha_{j \neq 0_0} = 0$.

\subsection{Evolution algorithm}

The dynamic noise term present in the Eq.~(\ref{SDE total with TBR hybrid 1})
means that we cannot directly apply the deterministic projected RK4IP
algorithm, which was used to obtain the results of \cite{Norrie2005b}. Rather
we must employ an algorithm that explicitly allows for such time-dependent
random processes.

The simplest such method is the Euler algorithm \cite{Press1992a}, in which the
drift terms are calculated at the start of each time step and the continuous
Wiener processes $dW$ are replaced by a single discrete Wiener process $\Delta
W$. However, for any reasonable accuracy the time step for an Euler algorithm
must be very small, thus requiring very long calculation times. Milstein and
Tretyakov \cite{Milstein2004a} have considered various more sophisticated
algorithms for propagating \SDE s with dynamic random processes. In particular,
they have shown that for situations where the influence of the dynamic noise on
the system is very much less than the deterministic evolution (the small noise
limit), one can accurately describe the total evolution using a relatively
simple modification to the fourth-order Runge-Kutta (RK4) algorithm.
Essentially, in this method one calculates the deterministic evolution using
the RK4 algorithm, while the dynamic noise is calculated using an Euler type
derivative calculation based on the state of the system at the start of each
time step. This result therefore allows us to use a slightly modified version
of the projected RK4IP algorithm to propagate Eq.~(\ref{SDE total with TBR
hybrid 1}).

Importantly, any numerical propagation method requires a discrete coordinate
space, such that the relations given for the spatial Wiener process,
Eq.~(\ref{spatial Wiener properties}, do not apply. Rather, we use noise sources
that obey
\begin{subequations}
\begin{eqnarray}
\left \langle dW_{\mu} \left( t \right) \right \rangle & = & 0 \\
\left \langle dW_{\mu} \left( t \right) dW_{\nu} \left( t \right)
\right \rangle & = & 0 \\
\left \langle dW_{\mu} \left( t \right) dW_{\nu}^* \left( t \right)
\right \rangle & = & \frac{1}{\Delta V} \delta_{\mu,\nu} \Delta t,
\end{eqnarray}
\end{subequations}
where $dW_{\mu} \left( t \right)$ is the time-dependent Wiener process at the
$\mu$th point on the coordinate space simulation grid and $\Delta V$ is the
volume space increment about that grid point. The algorithm advances the field
in time by the increment $\Delta t$ with each application, and we use $\Delta t
= 250$~ns for all our simulations.

\begin{figure}[t]
\begin{center}
\includegraphics{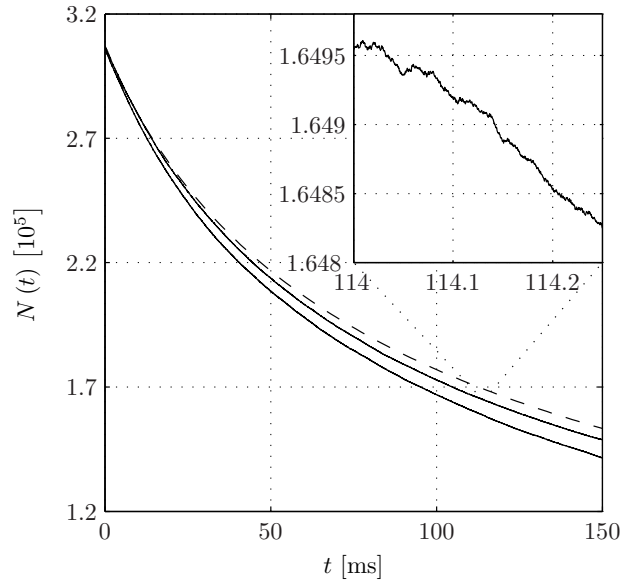}
\end{center}
\vspace{-0.5cm}
\caption{Total particle numbers for the sample trajectories of the system
described in the text. Main plot shows the populations for (higher and lower
solid lines) $v_{\rm cut} = \left\{ 44.3, 59.1 \right\}$~mms$^{-1}$, together
with the results for the simple model (dashed line). The inset shows the
population for the trajectory with $v_{\rm cut} = 44.3$~mms$^{-1}$ over a
subset of the total time. }
\label{Figure: TBR populations}
\end{figure}

The total number of real particles within the low-energy subspace is given by
Eq.~(\ref{Wigner field population}), where the subtraction of $1/2$ can be
understood as removing the virtual particles introduced into the initial state
of the field, Eq.~(\ref{trajectory initial state}). For systems with a large
number of modes $M$, an excellent estimate of the total particle number can be
made using
\begin{equation}
\label{Wigner total number 2}
N \left( t \right) \approx \sum_{j \in L} \left| \alpha_j \left( t \right)
\right|^2 - \frac{M}{2}.
\end{equation}
In Fig.~\ref{Figure: TBR populations} we plot the estimated total (real)
particle numbers calculated using Eq.~(\ref{Wigner total number 2}) for single
trajectories of the system described above using cutoffs of $v_{\rm cut} =
44.3$~mms$^{-1}$ and 59.1~mms$^{-1}$. From these curves we observe that, over
the larger time-scale, the total particle populations of the systems decrease,
apparently monotonically, with the $v_{\rm cut} = 59.1$~mms$^{-1}$ trajectory
showing greater particle loss. On the smaller time-scale however, as shown by
the inset, the total particle populations fluctuate rapidly, on a scale of
roughly 1--10 particles per time-step.

To provide a comparison with our truncated Wigner results, consider a simple
model. For a homogeneous system, and assuming that the third-order correlation
function is both spatially and temporally invariant, such that $g^{\left( 3
\right)} \left( {\bf x},t \right) = g^{\left( 3 \right)}$, Eq.~(\ref{general TBR
number loss}) can be integrated to return the time-dependent total particle
number
\begin{equation}
N \left( t \right) = \frac{N \left( 0 \right)}{\sqrt{1 + \frac{6 K_3 g^{\left(
3 \right)} N \left( 0 \right)^2}{V^2} t}}.
\end{equation}
For our system, using $g^{\left( 3 \right)} = 1$, the model shows a smaller
rate of loss than that observed for either of our trajectories, as shown by the
dashed line in Fig.~\ref{Figure: TBR populations}. This result is predicted by
Eq.~(\ref{expected rate of population change 2}), as is the difference in the
two trajectory populations.

\subsection{Results}

\section{Conclusion}

The truncated Wigner description of ultracold Bose gases has many significant
advantages over more traditional approaches, such as the \GPE, and the
extension outlined in this paper allows for the inclusion of three-body
recombination processes. We have shown that within the validity regime of the
Wigner truncation for two-body scattering, three-body recombination can be
described using stochastic differential equations describing the evolution of a
single trajectory, which can be solved using numerical techniques.

\begin{acknowledgments}
We wish to thank Dr. P.~B.~Blakie for informative discussions.
\end{acknowledgments}



\begin{widetext}
    \newpage
    \appendix

\section{Mathematical details}

Using the particular Wigner function given by Eq.~(\ref{Full Wigner factored})
in the full three-body recombination Wigner function equation of motion,
Eq.~(\ref{Wigner function evolution TBR full}), and evaluating all $\partial/
\partial \alpha_j, \partial / \partial \alpha_j^*$ operators returns

\begin{subequations}
\label{TBR Wigner evolution evaluated}
\begin{eqnarray}
\frac{\partial W}{\partial t}^{\left( {\rm TBR} \right)} \hspace{-0.5cm} & = &
\gamma \int d{\bf x} \left\{
- \left[ \left( \xi_{\mathcal{P}}^* - \xi_{\mathcal{P}_0}^* \right)
\Psi_{\mathcal{P}} + \Psi_{\mathcal{P}}^* \left( \xi_{\mathcal{P}} -
\xi_{\mathcal{P}_0} \right) \right]
\left( \left| \Psi_{\mathcal{P}} \right|^4 - 3 \left| \Psi_{\mathcal{P}}
\right|^2 \delta_{\mathcal{P}} + \frac{3}{2} \delta_{\mathcal{P}}^2 \right)
\right. \nonumber \\
&& \mbox{}
+ \left( 3 \left| \Psi_{\mathcal{P}} \right|^4
- 6 \left| \Psi_{\mathcal{P}} \right|^2 \delta_{\mathcal{P}}
+ \frac{3}{2} \delta_{\mathcal{P}}^2 \right) \delta_{\mathcal{P}}
\\
({\rm 2nd}) && \mbox{}
- \left[ \left( \xi_{\mathcal{P}}^* - \xi_{\mathcal{P}_0}^* \right)
\Psi_{\mathcal{P}} - \Psi_{\mathcal{P}}^* \left( \xi_{\mathcal{P}} -
\xi_{\mathcal{P}_0} \right) \right] \left( 6 \left| \Psi_{\mathcal{P}} \right|^2
- 6 \delta_{\mathcal{P}} \right) \delta_{\mathcal{P}}
+ \left( 6 \left| \Psi_{\mathcal{P}} \right|^2 - 3 \delta_{\mathcal{P}} \right)
\delta_{\mathcal{P}}^2 \nonumber \\
&& \mbox{}
+ \left( 6 \left| \Psi_{\mathcal{P}} \right|^4 - 12 \left| \Psi_{\mathcal{P}}
\right|^2 \delta_{\mathcal{P}} + 3 \delta_{\mathcal{P}}^2 \right) \left( \left|
\xi_{\mathcal{P}} - \xi_{\mathcal{P}_0} \right|^2 - \frac{1}{2} \sum_{j \in L}
\frac{\Gamma_j}{2} \left| \psi_j \right|^2 \right) \\
({\rm 3rd}) && \mbox{}
+ \frac{3}{2} \left[ \left( \xi_{\mathcal{P}}^* - \xi_{\mathcal{P}_0}^*
\right)^2 \Psi_{\mathcal{P}}^2 + \Psi_{\mathcal{P}}^{*2} \left(
\xi_{\mathcal{P}} - \xi_{\mathcal{P}_0} \right)^2 \right] \delta_{\mathcal{P}}
\nonumber \\
&& \mbox{}
- 3 \left[ \left( \xi_{\mathcal{P}}^* - \xi_{\mathcal{P}_0}^*
\right) \Psi_{\mathcal{P}} + \Psi_{\mathcal{P}}^* \left( \xi_{\mathcal{P}}
- \xi_{\mathcal{P}_0} \right) \right] \left[ \left| \Psi_{\mathcal{P}} \right|^2
\left( \left| \xi_{\mathcal{P}} - \xi_{\mathcal{P}_0} \right|^2 - \sum_{j \in L}
\frac{\Gamma_j}{2} \left| \psi_j \right|^2 \right)  \right. \nonumber \\
&& \mbox{} \left. \hspace{1cm}
+ \delta_{\mathcal{P}} \left(
\left| \xi_{\mathcal{P}} - \xi_{\mathcal{P}_0} \right|^2 - \frac{1}{2} \sum_{j
\in L} \frac{\Gamma_j}{2} \left| \psi_j \right|^2 - \frac{3}{2} \delta_P \right)
\right] \nonumber \\
&& \mbox{}
+ 6 \left| \Psi_{\mathcal{P}} \right|^2 \delta_{\mathcal{P}} \left( 2 \left|
\xi_{\mathcal{P}} - \xi_{\mathcal{P}_0} \right|^2 - \sum_{j \in L}
\frac{\Gamma_j}{2} \left| \psi_j \right|^2 \right) - 3 \delta_{\mathcal{P}}^2
\left( 2 \left| \xi_{\mathcal{P}} - \xi_{\mathcal{P}_0} \right|^2 + \frac{1}{2}
\sum_{j \in L} \frac{\Gamma_j}{2} \left| \psi_j \right|^2 - \frac{1}{2}
\delta_{\mathcal{P}} \right)
\\
({\rm 4th}) && \mbox{}
+ \left[ \left( \xi_{\mathcal{P}}^* - \xi_{\mathcal{P}_0}^*
\right)^2 \Psi_{\mathcal{P}}^2 + \Psi_{\mathcal{P}}^{*2} \left(
\xi_{\mathcal{P}} - \xi_{\mathcal{P}_0} \right)^2 \right] \left( 2 \left|
\xi_{\mathcal{P}} - \xi_{\mathcal{P}_0} \right|^2 - 3 \sum_{j \in L}
\frac{\Gamma_j}{2} \left| \psi_j \right|^2 \right) \nonumber \\
&& \mbox{} - \left[ \left( \xi_{\mathcal{P}}^* - \xi_{\mathcal{P}_0}^* \right)
\Psi_{\mathcal{P}} + \Psi_{\mathcal{P}}^* \left( \xi_{\mathcal{P}} -
\xi_{\mathcal{P}_0} \right) \right] \left( 6 \left| \xi_{\mathcal{P}} -
\xi_{\mathcal{P}_0} \right|^2 - 6 \sum_{j \in L} \frac{\Gamma_j}{2} \left|
\psi_j \right|^2 \right) \delta_{\mathcal{P}} \nonumber \\
&& \mbox{}
+ \left( 6 \left| \xi_{\mathcal{P}} - \xi_{\mathcal{P}_0} \right|^2 - 3
\sum_{j \in L} \frac{\Gamma_j}{2} \left| \psi_j \right|^2 \right)
\delta_{\mathcal{P}}^2 \\
({\rm 5th}) && \mbox{}
- \left[ \left( \xi_{\mathcal{P}}^* - \xi_{\mathcal{P}_0}^* \right)
\Psi_{\mathcal{P}} + \Psi_{\mathcal{P}}^* \left( \xi_{\mathcal{P}} -
\xi_{\mathcal{P}_0} \right) \right] \left( \left| \xi_{\mathcal{P}} -
\xi_{\mathcal{P}_0} \right|^2 - \sum_{j \in L} \frac{\Gamma_j}{2} \left| \psi_j
\right|^2 \right)^2 \nonumber \\
&& \mbox{}
+ \left[ 3 \left| \xi_{\mathcal{P}} - \xi_{\mathcal{P}_0} \right|^4 - 6 \left|
\xi_{\mathcal{P}} - \xi_{\mathcal{P}_0} \right|^2 \sum_{j \in L}
\frac{\Gamma_j}{2} \left| \psi_j \right|^2 + \frac{3}{2} \left( \sum_{j \in L}
\frac{\Gamma_j}{2} \left| \psi_j \right|^2 \right)^2 \right]
\delta_{\mathcal{P}}
\\
({\rm 6th}) && \mbox{}
+ \frac{3}{2} \left| \xi_{\mathcal{P}} - \xi_{\mathcal{P}_0} \right|^6
- 3 \left| \xi_{\mathcal{P}} - \xi_{\mathcal{P}_0} \right|^4
\sum_{j \in L} \frac{\Gamma_j}{2} \left| \psi_j \right|^2 \nonumber \\
&& \mbox{} \left.
+ 3 \left| \xi_{\mathcal{P}} - \xi_{\mathcal{P}_0} \right|^2
\left( \sum_{j \in L} \frac{\Gamma_j}{2} \left| \psi_j \right|^2 \right)^2
- \left( \sum_{j \in L} \frac{\Gamma_j}{2} \left| \psi_j \right|^2 \right)^3
\right\} W.
\end{eqnarray}
\end{subequations}
\end{widetext}
Here the terms arising from the separate derivative orders are kept separated,
as indicated on the left-hand side (first order derivative terms are given on
the first and second lines). For compactness we have written
$\delta_{\mathcal{P}} = \delta_{\mathcal{P}} \left( {\bf x,x} \right)$.


\begin{thebibliography}{10}

\bibitem{Moerdijk1996a}
A.~J. Moerdijk, H.~M. J.~M. Boesten, and B.~J. Verhaar, Phys. Rev. A {\bf 53},
  916  (1996).

\bibitem{Wieman1997a}
E.~A. Burt {\it et~al.}, Phys. Rev. Lett. {\bf 79},  337  (1997).

\bibitem{Dalibard1999a}
J. S{\"o}ding {\it et~al.}, App. Phys. B {\bf 69},  257  (1999).

\bibitem{Esry1999a}
B.~D. Esry, C.~H. Greene, and J.~P. {Burke, Jr.}, Phys. Rev. Lett. {\bf 83},
  1751  (1999).

\bibitem{Ketterle1995a}
K.~B. Davis {\it et~al.}, Phys. Rev. Lett. {\bf 75},  3969  (1995).

\bibitem{Norrie2005b}
A.~A. Norrie, R.~J. Ballagh, and C.~W. Gardiner, Submitted.

\bibitem{Jack2002a}
M.~W. Jack, Phys. Rev. Lett. {\bf 89},  140402  (2002).

\bibitem{Walls1994a}
D.~F. Walls and G.~J. Milburn, {\em Quantum Optics} (Springer-Verlag, Berlin,
  1994).

\bibitem{Gardiner2000a}
C.~W. Gardiner and P. Zoller, {\em Quantum Noise}, 3rd ed. (Springer-Verlag,
  Berlin, 2004).

\bibitem{Steel1998a}
M.~J. Steel {\it et~al.}, Phys. Rev. A {\bf 58},  4824  (1998).

\bibitem{Gardiner2003b}
C.~W. Gardiner, {\em Handbook of Stochastic Methods}, 3rd ed. (Springer-Verlag,
  Berlin, 2004).

\bibitem{Stringari2003a}
L. Pitaevskii and S. Stringari, {\em Bose-Einstein Condensation} (Oxford
  University Press, Oxford, 2003).

\bibitem{Fedichev1996a}
P.~O. Fedichev, M.~W. Reynolds, and G.~V. Shlyapnikov, Phys. Rev. Lett. {\bf
  77},  2921  (1996).

\bibitem{Ketterle1998a}
D.~M. Stamper-Kurn {\it et~al.}, Phys. Rev. Lett. {\bf 80},  2027  (1998).

\bibitem{Ketterle1999a}
J. Stenger {\it et~al.}, Phys. Rev. Lett. {\bf 82},  2422  (1999).

\bibitem{Press1992a}
W.~H. Press, S.~A. Teukolsky, W.~T. Vetterling, and B.~P. Flannery, {\em
  Numerical Recipes in C}, 2nd ed. (Cambridge University Press, Cambridge, MA,
  1992).

\bibitem{Milstein2004a}
G.~N. Milstein and M.~V. Tretyakov, {\em Stochastic Numerics for Mathematical
  Physics} (Springer-Verlag, Berlin, 2004).

\end{thebibliography}
\end{document}